\documentclass[aps,prd,amsmath,twocolumn,superscriptaddress,showpacs]{revtex4}
\usepackage{graphicx}
\usepackage{dcolumn}
\usepackage{bm}
\usepackage{natbib}
\usepackage{multirow}

\usepackage{hyperref}
\usepackage{ifthen}
\usepackage{amsmath}
\usepackage{amssymb}

\newcommand{\beq}{\begin{equation}}
\newcommand{\eeq}{\end{equation}}
\newcommand{\beqa}{\begin{eqnarray}}
\newcommand{\eeqa}{\end{eqnarray}}

\begin{document}

\title{Halo Zeldovich model and perturbation theory: \\ dark matter power spectrum and correlation function}  

\author{Uro\v s Seljak}
\email{useljak@berkeley.edu}
\affiliation{Physics, Astronomy Department, University of California, Berkeley, CA 94720, USA} 
\affiliation{Lawrence Berkeley National Laboratory, Berkeley, CA 94720, USA}

\author{Zvonimir Vlah}
\email{zvlah@stanford.edu}
\affiliation{Stanford Institute for Theoretical Physics and Department of Physics, Stanford University, Stanford, CA 94306, USA} 
\affiliation{Kavli Institute for Particle Astrophysics and Cosmology, SLAC and Stanford University, Menlo Park, CA 94025, USA}


\begin{abstract}

Perturbation theory for dark matter clustering has received a lot of attention in recent years, but its 
convergence properties remain poorly justified and there is no 
successful model that works both for correlation function and for power spectrum. 
Here we present Halo Zeldovich approach combined with perturbation theory (HZPT), in which we use
standard perturbation theory at one loop order (SPT)
at very low $k$, and connect it to a version of the halo model, for which we adopt 
the Zeldovich approximation plus a Pade expansion of a compensated one halo term. 
This low $k$ matching allows us to determine the one halo term amplitude and redshift evolution, 
both of which are in an excellent agreement with simulations, and approximately agree with the 
expected value from the halo model. 
Our Pade expansion approach of the one halo term added to Zeldovich approximation 
identifies a typical halo scale averaged over the halo mass function, the 
halo radius scale of order of 1Mpc/h, and a much larger halo mass compensation scale, which 
can be determined from SPT. 
The model gives better than one percent accurate predictions for the correlation function 
above 5Mpc/h at all redshifts, without any free parameters. 
With three fitted Pade expansion coefficients the agreement in power spectrum 
is good to a percent up to  $k \sim 1$h/Mpc,
which can be 
improved to arbitrary $k$ by adding higher order terms in Pade expansion. 

\end{abstract}


\maketitle

\setcounter{footnote}{0}

\section{Introduction}

One of the major puzzles in cosmology is understanding the nonlinear formation of structure in 
the universe. The current state of the art are the N-body simulations, which have been verified to 
give reliable answers at 1\% level in the power spectrum up to $k \sim 1$h/Mpc
\cite{2010ApJ...715..104H}, 
but are expensive to run, require large allocations, and are often not fully convergent using typical 
current generation box size and resolution. The convergence properties become a lot more difficult to achieve for higher
order correlations. 
For example, recent studies have shown that to reach one percent convergence on the covariance matrix one needs 
to simulate a volume in excess of 1000${\rm Gpc^3}$ \cite{2014arXiv1406.2713B}, and the covariance matrix depends both on the 
cosmological model and on the size of the survey one is simulating, making the numerical solution to the full problem 
an excessively demanding task given the current typical resources. 
Even more importantly, we want to use clustering statistics to extract information about our universe, and 
simulations do not provide much insight to the questions such as where is the information content and how 
to optimally extract it from the data. 

Alternative approach is to use perturbation theory (PT), 
of which the two most prominent examples are standard PT (SPT) and Lagrangian PT (LPT) (see \cite{2002PhR...367....1B} for a review). 
In this approach one assumes density perturbation is less than unity, and one expands the nonlinear equations 
perturbatively. Since density perturbations grow with wavevector $k$, this approach breaks down for some $k>k_{\rm nl}$. 
The current approaches typically underpredict power in LPT and overpredict in one loop 
SPT. 
For example, we know that Zeldovich approximation is quite successful in modeling 
the correlation function on large scales, including the baryonic acoustic oscillations (BAO) wiggles \cite{2014MNRAS.439.3630W}. 
However, it fails miserably in the 
power spectrum, underestimating the power at all but the lowest values of $k$, and giving predictions that 
are even below the linear theory at $z=0$ \cite{2015PhRvD..91b3508V}. Various one loop LPT extensions (one loop LPT \cite{2008PhRvD..77f3530M}, CLPT \cite{2013MNRAS.429.1674C}, CLPTs \cite{2015PhRvD..91b3508V}) 
somewhat improve this for the power spectrum, but make things worse for the correlation function \cite{2015PhRvD..91b3508V}. 
Physically, the problem of Zeldovich approximation and its extensions is that while it can describe
properly the initial streaming of dark matter particles, it fails to account for their capture into the dark 
matter halos: instead, the particles continue to stream along their trajectories, set by initial velocity 
in Zeldovich approximation, leading to an excessive smoothing of the power. 

Standard Eulerian PT (SPT) has the opposite problem. In power spectrum it quickly overpredicts the amount 
of power at one loop level, specially at low redshifts. The reason for this is that the loop integrals extend over all 
modes, including those that are in the nonlinear regime $k>k_{\rm nl}$. These modes are not in PT regime and 
typically these contrubutions are strongly suppressed 
in the simulations relative to PT, leading to too much power in SPT relative to simulations. 
Its Fourier transform, 
using a gaussian smoothing to obtain a  
convergent correlation function, results in a worse model than the Zeldovich approximation around BAO, but is otherwise comparable. 
Recent work emphasized this point in the context of effective field
theory (EFT) \cite{2012JHEP...09..082C}. 
For example, for power law power spectra these integrals can be divergent, so 
PT is wrong in such situations \cite{2013JCAP...08..037P}. 
Instead, it is argued that the best that one can do is to introduce effective field 
theory parameters that describe the correction from the small scale physics. 
For $k<1/R_h$ these terms can be expressed as a low $k$ limit of PT, but with free parameters. 
A priori it is unclear what is $R_h$, 
and how large these corrections are for our universe. In particular, realistic CDM power spectra
have the shape where most of the low $k$ limit loop integral comes from scales in the linear regime, where PT is 
believed to be valid, making the corrections 
from the nonlinear scales small, although probably not negligible. We will try to quantify this in more detail below. 
We will argue that halo model requires $R_h$ to be a typical halo radius. 

The philosophy we will advocate in this paper is that any analytic approach must 
give reliable results both in Fourier and in configuration space. 
Failure to do so is a sign of something 
missing in the model. For example, a Taylor series may give reliable results in Fourier 
space up to a certain $k$, but if truncated at a certain order it 
generally diverges at high $k$ and makes its Fourier transform imposible to calculate. 
The reason is that the series is not convergent at high $k$, and one has to adopt a 
different summation of the terms that have a better convergence. 
Our goal is to develop a model that is rooted in PT as much as possible, but is also 
able to reproduce simulations. Since simulations have been verified at 1\%, 
we will strive for this precision in this paper. 

\section{Halo Zeldovich model and Perturbation Theory} 

The main ingredients of our approach are the following: 

1) 1 loop SPT has loop integrals which, for low $k$ and high $z$, are entirely in the linear regime and thus reliably computed. 
There is no guarantee that the entire loop integral is correct at all redshifts, even for low $k$, 
but we will make this assumption here and derive the consequences. We will therefore assume that EFT corrections to 1 loop SPT 
are negligible at low $k$. For higher $k$ and low $z$ SPT 
predictions become increasingly unreliable and will not be used. Similarly, 2 loop SPT integrals are negligible at high $z$, 
while for low $z$ they extend deeply into the nonlinear regime and are grossly overestimated in SPT \cite{2015PhRvD..91b3508V}. 
Here we will simply ignore 2-loop SPT, with one exception, discussed next. 

2) Zeldovich approximation gives approximately correct physical picture of how the particles are displaced up to the process of halo formation, 
which stops the particles from displacing. The latter has very little effect on the Zeldovich displacement: most of the displacement is 
generated by modes in the linear regime and we will not be correcting Zeldovich approximation. In terms of SPT Zeldovich approximation 
receives contributions from loops at all order, but only from very specific terms related to the linear displacement field correlation function. 

3) Halo formation has to be an essential part of the complete model. Halos are objects of very high density, 
leading to a nearly white noise like contribution to the power spectrum at low $k$, with the halo profile parameters determining 
deviations from white noise at higher $k$. 
At high $k$, the halo term contribution dominates the correlations: all of the close pairs
are inside the same halos. At a scale $k \sim 1/2R_{\rm vir}$ the number of close pairs involves all of the 
pairs inside the virial radius, which must give a contribution of the other of $M^2$, where $M$ is the virial
mass of the halo. Integrating over all the halos, and weighting by the halo mass function $dn/dM$, one obtains an
estimate of one halo amplitude $\bar{\rho}^{-2}\int M^2 (dn/dM) dM$, 
where $\bar{\rho}$ is the mean density of the universe. At redshift 0 the integral is dominated by cluster mass halos with 
virial radius of 1-2${\rm h^{-1}Mpc}$. 
We expect the one halo term to be approximately of this amplitude at $k \sim 0.2-0.4{\rm h/Mpc}$. 
However, the mass has to be conserved so the halos have to be compensated, which forces the 
one halo term to vanish at very low $k$. There is no unique way to do this, since it depends on what 
we compensate against. The compensation is by definition a two halo term, since the mass is being compensated
by the particles outside the virial radius. 
Here we compensate against Zeldovich and
demand that the total agrees with SPT, which then automatically 
enforces mass and momentum conservation. 

4) More specifically, 
we will assume that 1) can be connected to 2) and 3) at some low $k$, which is low enough that SPT can be assumed to be valid, yet 
large enough to still be close to the scale which dominates the compensation. We will match the two on this scale 
where both descriptions are valid, and use 2)+3) at higher $k$. 
Since we believe that Zeldovich approximation is a good starting point for any modeling 
we will include it as one ingredient of the theory, and
decompose the power spectrum and correlation function into two parts,
\begin{equation}
P(k)=P_{\rm{Zel}}+P_{BB}, \,\,\,\, \xi(r) = \xi_{\rm{Zel}}(r)+\xi_{BB} (r). 
\end{equation}
Here subscript $Zel$ stands for Zeldovich
and $BB$ for broadband beyond Zeldovich \cite{2015PhRvD..91b3508V}, which is our one halo term. 
While there may be a residual BAO wiggle signature that is not 
captured by Zeldovich, it is essentially negligible in the power spectrum and at most a few percent in the
correlation function around BAO (100Mpc/h), probably too small to be observed by existing or future redshift surveys 
due to large sampling variance errors. 
Here we will thus focus on the modeling of the broadband one halo component and ignore the wiggle part.
Note that we do not assume that the Zeldovich part is uncorrelated with the one halo term. In this sense our one 
halo term is not the  so called stochastic term uncorrelated with Zeldovich. 

As we argued the key physics ingredient missing in PT is the halo formation, which leads to a large 
contribution from the near zero lag correlations, also called the one halo term in the halo model 
\cite{2000MNRAS.318..203S,2000MNRAS.318.1144P,2000ApJ...543..503M,2002PhR...372....1C}. 
The halo model postulates that the nonlinear evolution leads to halo formation, and that all the dark matter 
particles belong to collapsed halos, with the halo mass distribution given by the halo mass function $dn(M)$. 
The correlations between the dark matter particles can be simply split into correlations within the same halo, 
the one halo term, and between halos, the two halo term. On large scales the latter reduces to linear theory 
$P_L(k)$. 
The one halo term in the halo model has a simple physical interpretation on small scales, which is that all dark matter particles 
are inside the dark matter halos distributed with a radial halo density profile,
and this leads to a power spectrum that is simply an integral over the 
halo mass function times the Fourier transform of the halo profile squared. The halo profile has a compact support, 
extending out to roughly the virial radius, and its correlation function is a convolution of the 
profile with itself, extending to roughly twice that.  
In power spectrum the convolution becomes a square of the Fourier transform of the profile, which 
can be expanded as a series of even powers of $k$ \cite{2014MNRAS.445.3382M},
\begin{equation}
P_{BB}(k)=F(k)A_0\left(1-R_{1h,2}^2k^2+R_{1h,4}^4k^4+...\right). 
\end{equation}
Here the parameters $A_0R_{1h,n}$ have a specific interpretation in terms of the
integrals over the halo mass function $n(M)$ times halo mass $M$ squared, and times 
$2n$ moments of the halo radius averaged over the halo density profile \cite{2014MNRAS.445.3382M}. 
Specifically, 
$A_0=\bar{\rho}^{-2}\int M^2 dn(M)$ is just a weighted halo mass squared divided by the mean density $\bar{\rho}$ and does not 
depend on the halo density profile. These arguments however do not yet account for the halo compensation. 
Above we introduced $F(k)$, which
is the compensation function, required to vanish in $k \rightarrow 0$ limit as long as the two halo term converges
to linear theory in the same limit:
mass conservation requires that the leading nonlinear one halo term cannot be a constant $A_0$ \cite{2014MNRAS.445.3382M}.
Thus one halo term has to be generalized to include mass compensation effects: 
nonlinear effects cause the dark matter to collapse into dark matter halos, bringing in mass from larger scales, 
so it has to be compensated by a mass deficit at large scales  
to satisfy the mass (and momentum) conservation. Because of this 
one can show that the one halo term has to scale as $k^4$ at low $k$ 
\cite{1993ppc..book.....P,2000MNRAS.318..203S,2013PhRvD..87h3522V}. 

The two halo term can also be expressed as a convolution of the linear theory 
over the halo profiles, and the resulting Taylor 
expansion is given by 
a similar series 
\begin{equation}
P_{2h}(k)=P_L(k)\left(1-k^2R_{2h,2}^2+k^4R_{2h,4}^4...\right).
\end{equation}
The leading order correction scales as $k^2P_L(k)R_{2h,2}^2$, where $R_{2h,2}$ is also related to an 
average second moment of halo density profile, although with a different mass and halo bias 
weighting \cite{2000MNRAS.318..203S}. 
Note that this gives at the leading order correction the usual EFT term \cite{2012JHEP...09..082C}. 
It is clear that both one halo and two halo Taylor 
expansions break down for $k>R_{h,2}^{-1}$. The breakdown of the two halo term does not matter: at the relevant 
$k$ the correlations are dominated by the one halo term. For the latter however, a different expansion is 
advantageous, as we discuss below. 

In this paper we 
argue that the natural way to connect SPT 
to small scale nonlinear effects is in the context of the Zeldovich approximation plus a compensated one 
halo term. 
In this picture 
we can think of $P_{\rm{Zel}}$ as the leading order two halo term, and $P_{BB}$ as the one halo term. 
The motivation for this is that Zeldovich correlation function is almost exact for $r>5Mpc/h$, and that the 
correction relative to it is negative, suggesting a compensation of a nonlinear term $\xi_{BB}(r)$. 
All the corrections to the Zeldovich model thus go into the compensation term $F(k)$. 
In the halo model these corrections would arise from compensation of the halo term and from two halo term 
correlations of particles inside the halos. 
We do not try to separate 
these into the latter that one expects to scale as $k^2P_L(k)$ at low $k$, and the 
former that scales as $k^4$ at very low $k$. 
In general it is difficult to do this separation, as both of these
arise from the two halo term correlations. 
It is also not clear that in the regime where it matters ($k>0.1h/Mpc$) these low $k$ expansions still apply. 
We will however use them at very low $k \sim 0.02h/Mpc$, where we expect SPT to be valid. 
However, it should be clear that the term $P_{BB}$ is not just the 
one halo term in the traditional sense, because of these compensation corrections at low $k$. 

In \cite{2014MNRAS.445.3382M} this function $F(k)$ was simply fitted using a polynomial form. We will begin by 
keeping this function completely general, and then choosing a very simple form for it. 
Our one halo term, and thus the compensation form of $F(k)$, is defined relative to the 
Zeldovich term. If we adopted a different form for the two halo term we would obtain a different form of $F(k)$. 
In principle, Zeldovich approximation itself could contain some halos with correct mass and so a part of what 
we usually call one halo term could already be contained there. However, we will see that this 
must be a minor effect and the one halo term we derive agrees with the expected value from the halo model. 

While the one halo expansion in even powers of $k$ 
above works in Fourier space up to the virial radius scale of order $k \sim R_{1h,2}^{-1} \sim 1$h/Mpc, it breaks down above that.
Moreover, powers of $k$ diverge at high $k$ and 
do not have a well defined Fourier transform, making this form unsuitable for correlation 
function predictions without resumming it first. 
Instead we will use in this paper the Pade series ansatz
\begin{equation}
P_{BB}= A_0F(k){1+\sum_{m=1}^{n_{\rm max}-1} (kR_{m})^{2m} \over 1+\sum_{n=1}^{n_{\rm max}}(kR_{nh})^{2n}}.
\label{eq:pbb}
\end{equation}
By requiring the series in the denominator to run to a higher power than in the enumerator 
 we guarantee that the series does not diverge at high $k$ and has a finite Fourier transform. 
Here we will explore the truncation of the series at $n_{\rm max}=0, 1, 2$.
In terms of the halo model $A_0$ has the same interpretation as before, it is the
halo mass mass squared averaged over the halo mass function. 
It is the only quantity that has units of power spectrum, all the other parameters have 
units of length. 
For the $R_{m}$ and $R_{nh}$ parameters we expect that they will be related to a typical scale of the halos. 
There are several halo scales one can define: 
one is the scale radius $R_s$, defined as the scale where the slope of the density profile is -2. 
Another scale we can define is the virial radius $R_{\rm vir}=cR_s$, where $c$ is the concentration parameter with values around 
3-4 for the most massive halos and increasing towards less massive halos \cite{1997ApJ...490..493N}. 
The mean overdensity 
at $R_{\rm vir}$ is 200 by definition, while the typical mean overdensity at $R_s$ can be of order 1000 or more. 
In the halo model approach we integrate over these with the halo mass function, and the number of pairs 
for each is proportional to $M^2$, which gives most of the weight to the very massive halos. 
As a result,  
if we use $n_{\rm max}=1$ and only have one scale parameter, which we denote $R^0_{h1}$, then we 
find that the typical scale, 
when averaged over all halos, is of order 1Mpc/h at $z=0$, a typical scale of a cluster. 

While the halo model has been successful as a phenomenological model, its connection and consistency with PT has not been 
explored. 
In this paper we propose a Halo Zeldovich model applied to Perturbation Theory (HZPT) approach, 
in which we connect the halo model to PT in the regime where 
both can be expected to be approximately valid. 
We take the approach that one loop SPT has a regime of validity on very large scales and gives us the correct description of the 
onset of nonlinearity. This is not guaranteed by SPT: the one loop SPT may receive contributions from small scales which are 
nonlinear and thus not reliably computed. 
For CDM type power spectra, the integrals are convergent and for
sufficiently high redshift all of the one loop integral contributions come from linear scales, the prediction is reliable, and 
no EFT correction is needed. For now we will simply assume there are no corrections to SPT at low $k$, 
and return to this discussion later. 

Let us therefore assume that one loop SPT is correct at very low $k$, and that it can be matched to 
the halo model ansatz in equation in the regime of its validity. 
The low $k$ limit of Zeldovich 
approximation is \cite{2015PhRvD..91b3508V} $P_{\rm Zel}=(1-k^2\sigma_{L}^2+k^4\sigma_{L}^4/2)P_L +Q_3/2$, where $P_L$ is the linear power spectrum, 
$\sigma_L^2=1/(6\pi^2) \int dq P(q) $ is the square of linear displacement 
field dispersion and $Q_3=1/(10\pi^2) k^4 \int dq P^2(q) / q^2=C_3k^4$ is a mode coupling integral, as defined in \cite{2008PhRvD..77f3530M}. 
It has been shown in \cite{2015PhRvD..91b3508V} that this expansion is valid for $k<0.1$h/Mpc. Note that we kept terms beyond 1-loop in the 
Zeldovich. 
One loop SPT can be written as $P_{\rm SPT}=P_L+P_{13}+P_{22}$, which 
at low $k$ is $P_{13}=-61/105k^2\sigma_{L}^2P_L$ and $P_{22}=45Q_3/98$. 

\begin{figure*}[tb]
\includegraphics[scale=0.62 ]{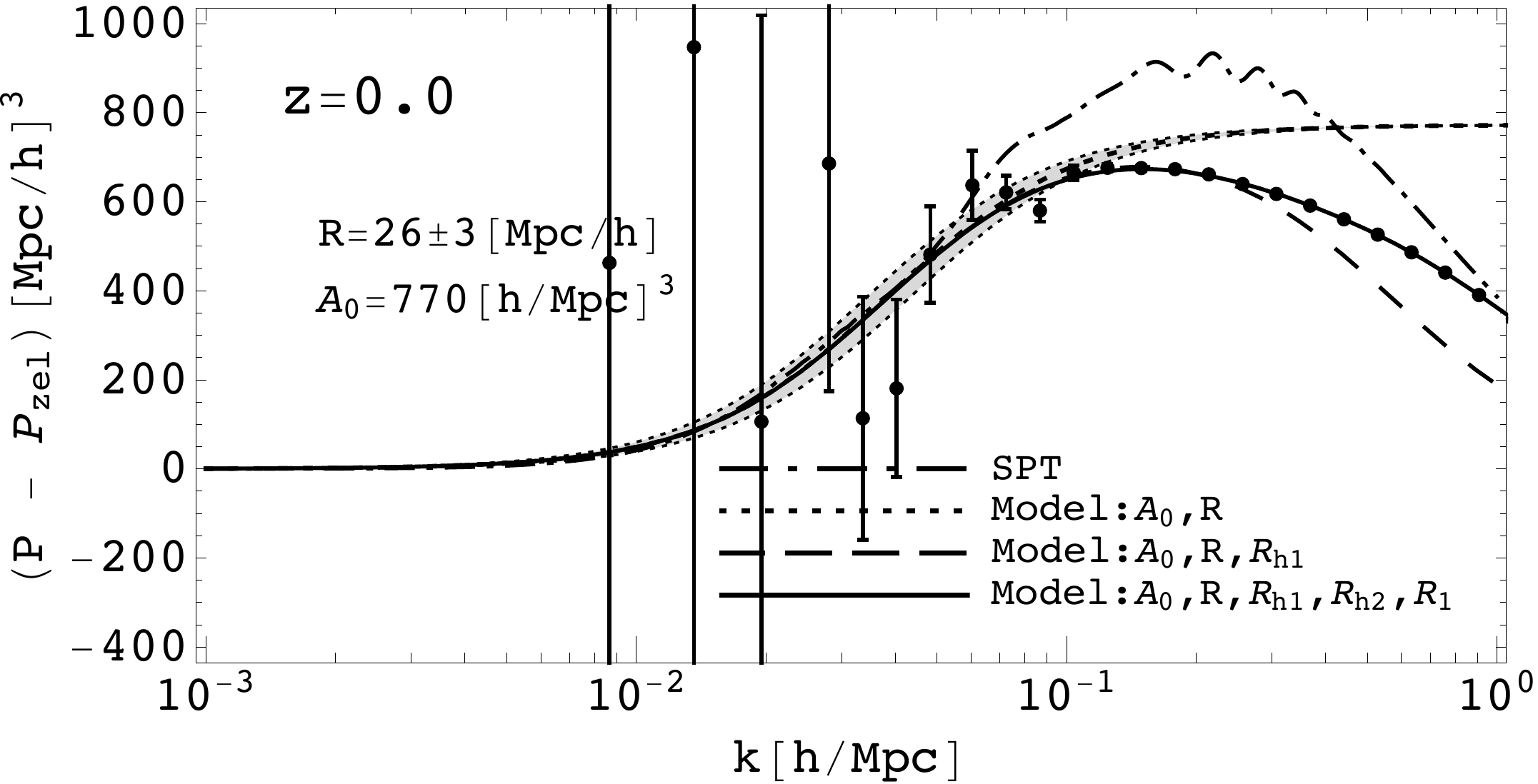}
\caption{HZPT model for $n_{\rm max}=0,1,2$ (dotted, dashed and solid lines respectively), 
together with SPT (dot-dashed line) are shown as a function of $k$. 
We match SPT to HZPT around $ k \sim 0.02$h/Mpc to derive $A_0$ and $R$, while the 
remaining parameters of our model ($R_1$, $R_{1h}$, $R_{2h}$) are fitted to simulations (points) at scales $ k > 0.1$h/Mpc. 
Gray band shows the change in the HZPT, $n_{\rm max}=0$ model in case where $R$ value varies for $\pm 3$Mpc/h.
This plot is for $z=0$, and higher $z$ versions are similar. Simulation points are taken from \cite{2015PhRvD..91b3508V}, and are 
for $\sigma_8=0.807$ Lambda CDM model. 
}
\label{fig1}
\end{figure*}

Let us begin by first dropping the 2-loop term $k^4\sigma_{L}^4P_L/2$ from Zeldovich approximation. 
Then all of the one loop terms scale as the square of the power spectrum. 
Matching at low $k$ gives
\begin{equation}
P_{\rm SPT}(k)-P_{\rm{Zel}}(k)={44 \over 105}k^2\sigma_{L}^2P_L-{2 \over 49}Q_3=A_0 F(k), 
\label{aor}
\end{equation}
where the linear order cancels in the difference and we have dropped all higher order terms since they are negligible at low $k$. 
What does this imply for the amplitude dependence of $A_0$ and $F(k)$? 
We will adopt the standard $\sigma_8$ normalization for the amplitude of fluctuations, where 
$\sigma_8(z)$ is the rms fluctuation of spheres of radius of 8Mpc/h, and which is redshift dependent.
Often this is phrased in terms 
of the redshift dependence of the growth factor $D(z)$, and in linear theory we would write $\sigma_8(z)=D(z)\sigma_8(z=0)$. 
But the amplitude dependence is more general than the redshift dependence, since it encompases the idea that changing 
the redshift or changing the amplitude should give the same result in the context of PT. 
Since both SPT and Zeldovich at low $k$ scale as a square of the power spectrum, which itself scales as $\sigma_8^2$,
requiring equation \ref{aor} to be valid over a broad range of $k$ where $F(k)$ is rapidly changing, there can only 
be one solution to the amplitude dependence,
$A_0 \propto \sigma_8^4$ and $F(k)={\rm const}$. This simple result is in very close agreement with 
simulations spanning a wide range of redshifts and models \cite{2014MNRAS.445.3382M}, where the slope of 3.9 was derived. 
We will see below that including the full Zeldovich instead of its lowest order further improves the agreement on the slope. 
Note that this is valid for a general compensation function $F(k)$. 

To proceed we need to assume a specific functional form for the compensation term. 
A simple way to achieve compensation is to use $F(k)=1-1/(1+k^2R^2)$, which has an analytic Fourier transform and 
vanishes as $k \rightarrow 0$, so that the overall model for one halo term is  
\begin{equation}
P_{BB}= A_0\left(1-{1 \over 1+k^2R^2}\right){1+\sum_{m=1}^{n_{\rm max}-1} (kR_{m})^{2m} \over 1+\sum_{n=1}^{n_{\rm max}}(kR_{nh})^{2n}}.
\label{eq:pbb}
\end{equation}
The parameter $R$ governs the transition of the one halo term to 0 at low $k$: this is the compensation scale parameter, and we 
expect it to be very large compared to the typical halo size. 
Here we simply chose the simplest form for this function $F(k)=(1-1/(1+k^2R^2)$,
but we do not expect this to be the correct form at all $k$, since, as seen from equation \ref{aor}, the residual between SPT and Zeldovich at 
low $k$ is not $k^2$, but $k^2P_L$. 
We evaluate SPT and Zeldovich in the low $k$ limit and fit for parameters $A_0$ and $R$. We find there is a 
range of $k$ where the fit is good (figure \ref{fig1}) and we use $0.02{\rm h/Mpc}<k<0.05{\rm h/Mpc}$ for the fits. 
The fit is not perfect, a consequence of the simplified form of $F(k)$ term, but is a good fit for $0.02{\rm h/Mpc}<k<0.05{\rm h/Mpc}$. 
Even for the more general forms of $F(k)$ one can still define a typical compensation 
scale on which $F(k)$ goes from unity to zero. Current form of compensation is sufficient for our purposes and we will not explore 
more general forms. 
We find
\begin{eqnarray}
A_0 &=& 750\left({\sigma_8(z) \over 0.8}\right)^{3.75} \text{(h/Mpc)}^3 \nonumber\\
R &=& 26.0 \left({\sigma_8(z) \over 0.8}\right)^{0.15} \text{(Mpc/h)}
\label{a0r}
\end{eqnarray}
This is a remarkably simple result. Moreover, it agrees well 
with the recent numerical determination of $A_0$ amplitude from a suite of 38 emulator simulations 
at different redshifts in \cite{2014MNRAS.445.3382M}, where a scaling
$A_0=670(\sigma_8/0.8)^{3.9}({\rm Mpc/h})^3$ has been derived over the redshift range 0-1. The amplitude is a bit different 
because the form of compensation used in this letter is a bit different than in \cite{2014MNRAS.445.3382M}, but they 
both provide equally good fit to the simulations, as shown in figure \ref{fig1}. 
The two parameters are correlated, and parameter $R$ is less well determined than $A_0$: $R$ comes with five times
larger relative error than the error on $A_0$. Figure \ref{fig1} shows the band over which $R$ is varied by 12\%: a reasonable 
estimate of its error is 3-5\% (and the error on $A_0$ is 1\% or less).  We find the slopes of $A_0$ and $R$ to be uncertain at 0.1 level. 
The value of the amplitude also agrees very well with the expected amplitude of the one halo term in the halo model, 
which is $\bar{\rho}^{-2}\int M^2 dn$, where $M$ is the halo mass, $dn/dM$ is the halo mass function and 
$\bar{\rho}$ is the mean density of the universe. 
This suggests that Zeldovich approximation on itself does not contribute much to the one halo term. 

To qualitatively derive a value for $R$ and $A_0$ let us look at the low $k$ limit of SPT, focusing on the 
leading order $k^2$ and $k^4$ terms at the peak of the power spectrum around $k_{\rm peak} \sim 0.02$h/Mpc, 
where $P_L(k_{\rm peak}) \sim 27000(\sigma_8/0.8)^2{\rm (h/Mpc)^3}$ 
and $\sigma_L^2 \sim 36(\sigma_8/0.8)^2({\rm Mpc/h})^2$. Since the difference between Zeldovich and SPT for the 
low $k$ gives $(44/105)k^2\sigma_L^2P_L$, by equating that to $k^2A_0R^2$ (low $k$ limit of equation \ref{aor}) 
we derive $R \sim (0.42\sigma_L^2P_L(k_{\rm peak})/A_0)^{1/2}
\sim  24$Mpc/h for the best fit value of $A_0$, in a qualitatively 
good agreement with the value of 26Mpc/h derived numerically. 
Thus the value of $R$ is determined by the linear power 
spectrum amplitude at the peak, rms displacement field and the amplitude of the one halo term $A_0$. 
To determine both $A_0$ and $R$ we need to expand SPT and Zeldovich up to $k^4$ around $P_L(k_{\rm peak})$, for which 
we also need to numerically evaluate $C_3 \sim 2\times 10^9{\rm Mpc/h}$. Matching $k^4$ terms gives 
$[(1/2-45/98)C_3+(\sigma_L^2)^2P_L/2]=A_0R^4$, 
which, when combined with $k^2$ term gives $R \sim 20$Mpc/h and $A_0 \sim 700({\rm Mpc/h})^3$ at $z=0$. 
We note further that there is a considerable contribution beyond $k^4$ from $P_{13}$ already around $k \sim 0.02{\rm h/Mpc}$, and 
in the numerical fits there is some correlation between $A_0$ and $R$. 

The Zeldovich term 
beyond 1-loop, $(\sigma_L^2)^2P_L/2$, cannot be neglected compared to the rest of $k^4$ terms even for $k<0.05{\rm h/Mpc}$, and 
as a consequence the scaling of $A_0 \propto \sigma_8^4$ and $R={\rm const}$ is mildly broken. This term is larger for low redshifts, 
explaining why the slope of $\sigma_8$ scaling is less than 4.0 and closer to 3.8. Remarkably, that is exactly what simulations suggest. 
Given the uncertainties in the form of $F(k)$ we cannot address in detail the remarkably small scatter of $A_0$ 
against the amplitude $\sigma_{11.3}$ when the shape of the power spectrum is varied \cite{2014MNRAS.445.3382M}. 

The leading order for our compensation term of the one halo $F(k)$ is $k^2$. It is often stated that 
the mass and momentum conservation effects generate $k^4$ tail \cite{2008PhRvD..77b3533C}, and so one would naively expect 
the leading term of one halo to be $k^4$. This term is generated by $P_{22}$ in SPT.
However, nonlinear evolution also leads to propagator effects contained in $P_{13}$, which at the leading order 
give $k^2P_L$ corrections to the linear theory and we have seen that these effects dominate at low $k$. 
We have assumed the two halo term to be Zeldovich approximation, which contains part of the $k^2P_L$ term in SPT but not all, 
and so the difference is still given by $k^2P_L$ term at the leading order, which has to be the leading order one halo term, 
and happens to coincide with $k^2$ at the peak of the power spectrum around $k \sim 0.02$h/Mpc, where we fit to our ansatz. 
As discussed above, we could use a more general form of $F(k)$ that would make the compensation and two halo terms 
exact, but we found this makes 
no practical difference to the final results. 
It is also possible to make a different two halo ansatz where $k^2P_L$ is entirely cancelled, for example, our two halo ansatz could be 
simply $P_{13}$ of SPT, or its nonlinear propagator version \cite{2008PhRvD..77b3533C}. 
As shown in \cite{2008PhRvD..77b3533C}, this ansatz leads to one halo term that is several times larger than expected in the 
halo model. 
It is possible to generate a valid halo model based on a different ansatz for the two halo term, but we do not pursue this further here. 


\section{Power spectrum predictions of HZPT}

In this section we fit higher order parameters of the one halo term to obtain the best possible agreement against simulations. 
The best fits for these parameters as powers of amplitude $\sigma_8(z)$ give
\begin{eqnarray}
R^0_{1h}&=&  1.87{\rm Mpc/h}\left({\sigma_8(z) \over 0.8}\right)^{-0.47},\nonumber\\
R_{1h} &=& 3.87{\rm Mpc/h}\left({\sigma_8(z) \over 0.8}\right)^{0.29},\nonumber\\
R_{1} &=& 3.33{\rm Mpc/h}{\sigma_8(z) \over 0.8}^{0.88},\nonumber\\
R_{2h} &=& 1.69{\rm Mpc/h}\left({\sigma_8(z) \over 0.8}\right)^{0.43}.
\end{eqnarray}
Here $R^0_{1h}$ refers to $n_{\rm max}=1$ case, while $R_{1h}$, $R_{2h}$ and $R_{1}$ refer to $n_{\rm max}=2$ case. 
Even though we only fit to one set of simulations, 
we expect that these parameters are nearly universal and apply well to all cosmological models, just as in the case of the 
halo plus Zeldovich model of \cite{2014MNRAS.445.3382M}. The main difference relative to \cite{2014MNRAS.445.3382M} is the form of the 
compensation function $F(k)$, which was fitted to a 10-th order polynomial in \cite{2014MNRAS.445.3382M}, while here we adopt a 
much simpler form of equation \ref{eq:pbb}, and the expansion of the one halo term, which was in even powers of $k$ in 
\cite{2014MNRAS.445.3382M}, while we use Pade expansion here. 

\begin{figure*}[tb]
\includegraphics[scale=0.50]{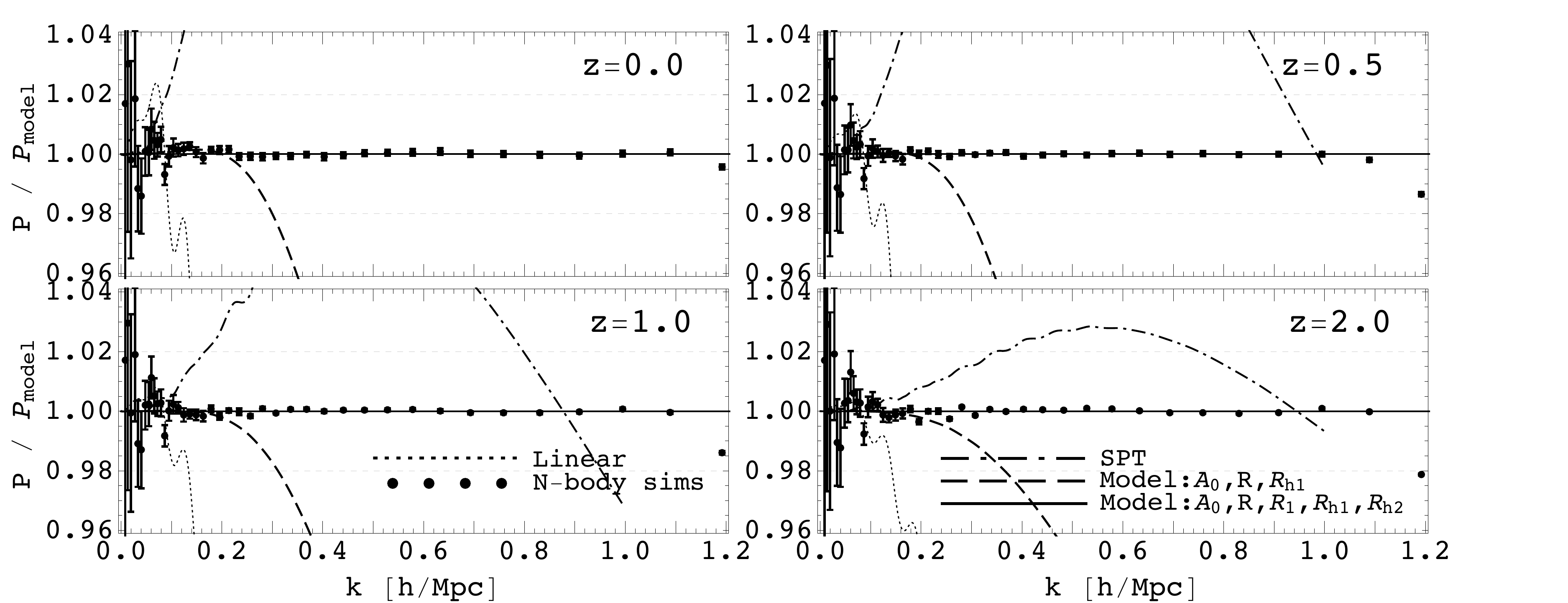}
\caption{Simulation power spectrum (points) is shown relative to HZPT model prediction ($P_{\rm model}$) with $n_{\rm max}=2$ (black solid line)
at four different redshifts ($z=0.0$, $0.5$, $1.0$ and $2.0$). 
Also shown are SPT (dot-dashed line), HZPT with $n_{\rm max}=1$ (dashed line) and linear theory (dotted line). 
Simulation points are taken from \cite{2015PhRvD..91b3508V}, and are
for $\sigma_8=0.807$ Lambda CDM model.}
\label{fig:PBBplot}
\end{figure*}

In figure \ref{fig:PBBplot} we show the HZPT for $P(k)$ against 
the N-body simulations for four different redshifts. We find that the model with $n_{\rm max}=1$
can fit the simulations at 1\% up to $k \sim 0.3$h/Mpc, while 
for $n_{\rm max}=2$ the fits are good to 1\% up to $k \sim 1$h/Mpc. 
Parameter $R$ ensures the low k behaviour of the model, while $A_0$ sets the
peak amplitude of $P_{BB}$ which is around $k\sim 0.12$h/Mpc for all redshifts. 
In \cite{2006ApJ...651..619J} has been argued that 
SPT is a relatively good description against N-body simulations for $z>3$. This is consistent 
with our results: as shown in figure \ref{fig:PBBplot}, for $z=2$ SPT differs from simulations by only 2\%, 
and this difference is presumably even smaller for higher $z$. 

We have established that one loop SPT can determine $A_0$ and $R$. 
Which other coefficients of equation \ref{eq:pbb} can PT determine? Since coefficients $R_{nh}$, $R_n$, $n>0$, 
are determined by a typical halo scale averaged over the halo profile and averaged over the halo mass function, 
they depend on the regime where the overdensity is very large:
halo virial radius $R_{\rm vir}$ is defined at a mean overdensity of 200, and $R_s$ is even smaller 
(with the correspondingly larger mean overdensity). 
These cannot be computed from standard PT, which is only supposed to work in the regime where $\delta<1$. 
One can show (figure \ref{fig:PBBplot}) that these terms become important at a percent level
around $k \sim 0.2$h/Mpc at $z=0$. It seems unlikely that PT can make predictions at a percent level for $k>0.2{\rm h/Mpc}$, 
regardless of which PT formalism we use.
It is however remarkable that one loop SPT can predict the amplitude of the one halo term $A_0$. 
This is possible because $A_0$ depends on the total halo 
mass that has nonlinearly collapsed and not on its profile. 
At low $k$ the non-perturbative halo profile effects also give rise to the two halo correction term of order 
$k^2R_{2h}^2P_L$, where $R_{2h} \sim 1$Mpc/h. This is the EFT term of \cite{2012JHEP...09..082C}. 
At low $k$ this term is a small, a few percent, correction to 
the SPT term, which is of course small compared to linear theory. At higher $k$ this term is modified and in 
our model absorbed into the overall compensated one halo term $P_{BB}$. 

\section{Correlation function predictions of HZPT}

We would like to require from a good PT model that it 
works both in Fourier and in configuration space, but so far there has been 
no successful model achieving this.
Typically, Zeldovich approximation works quite well in correlation function but fails in the power 
spectrum, while SPT does not give very good correlation function predictions, specially around BAO. 
In HZPT approach, we expect Zeldovich term to dominate the correlation function at large radii. 
In the absence of compensation the one halo term would be limited to scales around twice the virial radius and below. 
With compensation these effects extend to large radii, but as we will show, remain small. 
There is thus a crucial difference of the effect of the one halo term 
between correlation function and power spectrum: the one halo term is mostly a few percent effect 
for $r>5{\rm Mpc/h}$ in correlation function, caused by compensation effects. 
On the other hand, one halo term  
can be very large for $k>1/R\sim 0.04{\rm h/Mpc}$ in the power spectrum and dominates $P(k)$ for $k>0.2{\rm h/Mpc}$. 

\begin{figure*}[tb]
\includegraphics[scale=0.495]{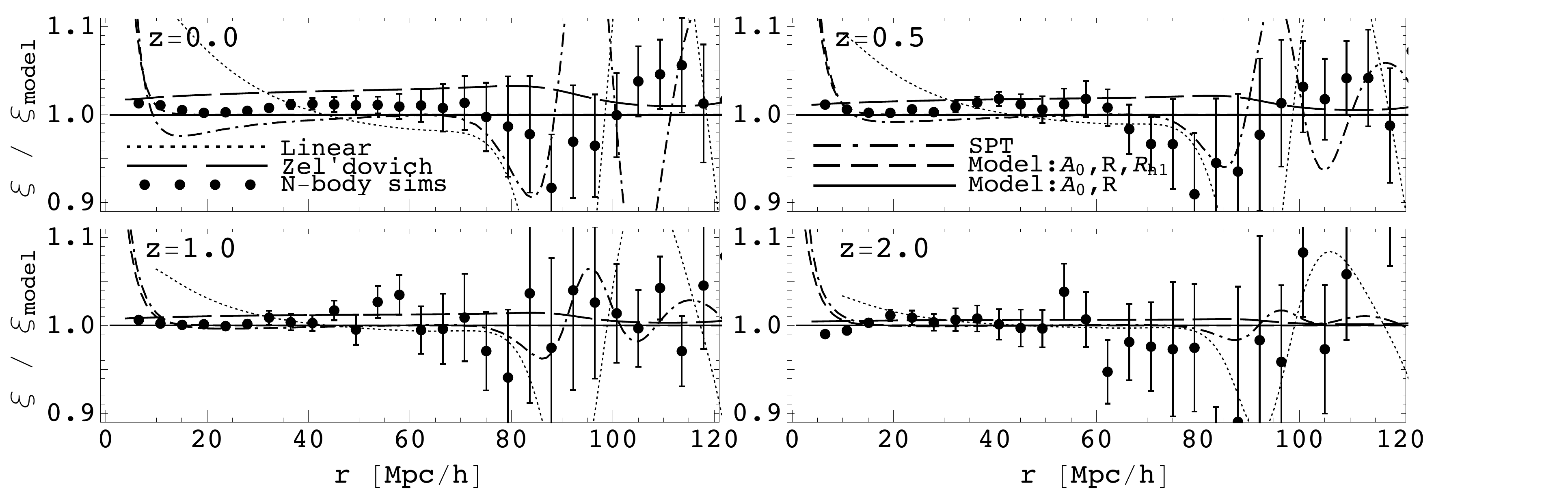}
\caption{Simulation correlation function (points) is shown relative to HZPT model prediction ($\xi_{\rm model}$) with $n_{\rm max}=0$ 
(black solid line) at four different redshifts ($z=0.0$, $0.5$, $1.0$ and $2.0$). Note that HZPT $\xi_{\rm model}$ with $n_{\rm max}=0$ 
if completely determined with the $A_0$ and $R$ parameters obtained from the low $k$ of SPT power spectrum. 
Also shown are HZPT with $n_{\rm max}=1$ (dashed line), Zeldovich (long dashed line), SPT (dot-dashed line), and linear theory (dotted line).
Simulation points are taken from \cite{2015PhRvD..91b3508V}, and are
for $\sigma_8=0.807$ Lambda CDM model.
} 
\label{fig:xi}
\end{figure*}

An advantage of the ansatz in equation \ref{eq:pbb} is the existence of analytical Fourier transform for low $n_{\rm max}$. 
On scales $r \gg R_{\rm halo}$, and assuming $R \gg R_{nh},R_n$, one can use $n_{\rm max}=0$ and find
\begin{equation}
\xi_{BB}(r)= - \frac{A_0e^{-\frac{r}{R}}}{4\pi rR^2}.
\label{xibb}
\end{equation}
Similarly, if we keep the leading $R_{h1}$ effects in $n_{\rm max}=1$ case we get:
\begin{equation}
\xi_{BB}(r)= - \frac{A_0e^{-\frac{r}{R}}}{4\pi rR^2} \left(1 - \left(\frac{R}{R_{h1}}\right)^2 \exp \left[-\frac{R + R_{h1}}{R R_{h1}} r\right] \right).
\label{xibb_v2}
\end{equation}
First equation effectively reduces the low $k$
model that we started with in Fourier space to a model in the
configuration space with the same two parameters, $A_0$ and $R$, that
have been determined using SPT. 
Since the virial radius of the largest halos  
is about 2Mpc/h at $z=0$, we expect the transition between the two regimes to be around 4Mpc/h. 
The results for the correlation function are shown in figure \ref{fig:xi}. 
We see that our model significantly improves upon other PT results, achieving
1-2\% agreement down to 5Mpc/h at all redshifts. 
A possible exception is the BAO wiggle, r$>$80Mpc/h, where there may be additional wiggle contribution 
that was discussed in \cite{2015PhRvD..91b3508V}, but it is unclear whether it is real 
given the large sampling variance fluctuations. 
It also improves upon Zeldovich approximation, which 
is already very good by itself, with only a few percent deviations from simulations over this range. 
Our model reduces the correlation function relative to Zeldovich, as expected by the compensation
effects, which take mass from large scales to enhance one halo terms on small scales. 
As expected our resuts agree with SPT on large scales, but only away from BAO, making the range where SPT agrees 
with simulations and our model at 1\% level only around 50-70Mpc/h. 

Given that the one halo contribution relative to Zeldovich is a few percent only, 
any further corrections 
have a very small effect on $\xi(r)$ for $r>5{\rm Mpc/h}$. 
For example, one can improve the agreement of the model with simulations somewhat by increasing $R$. 
We can do this at a few percent level since this is the formal error from the fits to SPT. 
In the context of our approach, increasing $R$ beyond its SPT predicted value at a level more than this
would only be possible if there are nonperturbative or 2-loop corrections at low $k$, since
an assumption of our model is that one loop SPT is the correct theory at low $k$.
Note that EFT corrections as advocated by \cite{2012JHEP...09..082C} will reduce $A_0R^2$, which goes in the opposite direction. 
Nevertheless, as discussed above, there is no guarantee for SPT to be true even for very low $k$: the loop integrals 
can extend into the regime where density perturbation exceeds unity. 
At $z=0$, for CDM models, this happens 
around $k_{\rm nl} \sim 0.1-0.2$h/Mpc. The low $k$ expansion of SPT is dominated by $k^2\sigma_L^2P_{\rm lin}(k)$ terms, which 
depend only on the convergence of the $\sigma_L^2$ integrals. This
converges to about 90\% of the value for $q\lesssim k_{\rm nl} \sim 0.2$h/Mpc at $z=0$, 
and converges even more to its full value for $z>0$ where $k_{\rm nl} > 0.2$h/Mpc \cite{2015PhRvD..91b3508V}. So we expect the 
one loop SPT to be almost perfectly valid at low $k$ for high redshifts, but there may be low redshift corrections 
at a several percent level, which
may allow for additional changes in $R$ or $A_0$ beyond the predicted value at at comparable level. 
In addition, our form of the compensation term is just an assumed ansatz, which could be modified for a better agreement. 

A related question is whether one should include higher loop contributions. 
At a next order in PT is the two loop SPT or, similarly (although not equivalently), one loop LPT. 
In both cases the corrections to one loop SPT can become important at low redshifts, 
even at low $k$. For example, at $z=0$ the two loop SPT correction to $P_{13}$ is about 15\% \cite{2015PhRvD..91b3508V}, 
and since the relative correction of two loop to one loop scales as $\sigma^2_8$, 
the correction is much smaller at higher redshifts. 
However, higher loop contributions are likely to be 
grossly overestimated in PT. For example, comparison against simulations suggests that the one loop LPT 
contribution to rms displacement is in reality almost entirely suppressed, such that the total 
nonlinear value of $\sigma_{NL}^2$ differs by only 1-2\% relative to the 
linear value $\sigma_L^2$ at all redshifts \cite{2014PhRvD..89h3515C,2014JCAP...06..008T,2015PhRvD..91b3508V}. 
Physically this can be understood by the process of halo formation, which stops particles from 
displacing on small scales, and instead traps them inside the dark matter halos: the large scale displacements, 
which are one loop in SPT, are correctly predicted, 
while the small scale displacements, two loop and higher in SPT sense, are strongly suppressed. 
It thus seems better to drop  two loop terms entirely, although we have no formal proof of this statement. 

In summary, formally one cannot exclude corrections at low $k$, which will be of the EFT form $k^2P_L$, 
but these are likely to be of order 
a few percent only. We see no need for such corrections in our approach: 
our model is accurate at the current precision of simulations. 
We thus argue that one loop SPT is close to the correct theory for 
$k < R^{-1}$, but this is not a result that can be formally derived. 

\section{Conclusions} 

In this paper we develop a model for dark matter power spectrum and correlation function that 
is 1\% accurate for both, and that is based on perturbation theory (PT) as much as possible. We argue that PT 
approaches to large scale structure can only have a hope of being valid for very large scales, $k<0.05{\rm h/Mpc}$, a regime that
we usually do not focus on when comparing PT to simulations, since deviations from linear theoryare very small there. 
We also argue that Zeldovich approximation is a useful starting point for any halo based model, and that halo formation 
has to be essential part of the model. 
We propose a model which we call Halo Zeldovich PT (HZPT) model, in which 
Zeldovich approximation is supplemented with the one halo term, and the sum of the two is connected to one loop
standard PT at low $k$. 
Within this model we derive the one halo term amplitude $A_0 =750{\rm (Mpc/h)}^3\sigma_8^{3.8}$, which agrees with simulations 
both in amplitude and in $\sigma_8$ scaling. 

The one halo term needs to be compensated by the other halos for the mass conservation, and there are nonlinear 
contributions from two halo correlations, both of which we model this using a very simple functional form. 
This compensation scale of the one halo term has
effects on the power spectrum at a 
percent level or smaller, but we have argued it is essential in order to have a 
self-consistent model that connects the halo model to PT:
its introduction gave us 
one percent accuracy on both the power spectrum and the correlation function. 
In particular, the deviations of the correlation function of simulations from Zeldovich is negative and a few percent only, 
and this term explains its origin. 
We have argued that the regime where SPT is valid in the power spectrum, 
is at best limited to $k \sim 1/R \sim 0.04{\rm h/Mpc}$, while for higher $k$ one halo term (generalized by the 
two halo term corrections encoded in $F(k)$)
begin to dominate. It is not possible to 
formally exclude presence of nonperturbative or higher loop correction terms even at 
very low $k$, but we see no need to consider 
them and they are likely at most several percent for our universe. 
These terms will also generate a correction to the BAO wiggles \cite{2015PhRvD..91b3508V}
that we have ignored in 
this paper. 

We have proposed a Pade type expansion of the one halo term as a useful functional form that allows one to 
go beyond the convergence radius of the Taylor expansion. We have argued that the value of this radius is 
around 1Mpc/h, a typical virial radius of halos properly averaged over the halo mass function, 
hence Pade expansion is necessary if one wants a valid description for $k>1$h/Mpc. 
Our approach is similar to the treatment of nonlinear redshift space distortions (the so called Fingers of God, FoG), 
which also require to have an expression valid for $k>0.2$h/Mpc  
(the FoG scale is typically 5Mpc/h). In the context of FoG a Lorentzian distribution is often used, which is 
just the Pade series at the first order. 
Pade expansion also has the advantage of 
being convergent in both the power spectrum and the correlation function, making the Fourier 
transforms calculable. With this expansion, and keeping terms up to 2nd order, we are able to 
match power spectrum to better than 1\% against simulations, up to $k=1{\rm h/Mpc}$. 
The correlation functions also agree against simulations to this accuracy down to 5Mpc/h. 
We expect that a Pade ansatz for one halo term will be useful in modeling other correlation functions as well, such as 
galaxy-dark matter and galaxy-galaxy correlations. 
In principle a Pade expansion would also be needed for the two halo terms that scale as $P_L$, but in practice the
two halo term is irrelevant on scales where this would make any difference ($k \sim 1$h/Mpc), 
so a simple Taylor expansion giving rise to $k^2P_L...$ terms suffices.  

We have argued that the predictive power of PT beyond Zeldovich has been reduced to two numbers, $A_0$ and $R$. 
To improve the model further one needs to provide information on the 
halo profiles in the deeply nonlinear regime, 
which is unlikely to be predictable by the PT. It has been argued in \cite{2014MNRAS.445.3382M} 
that these coefficients are also not predictable by 
N-body simulations, due to the baryonic effects, 
so PT is not necessarily inferior to N-body simulations. 
For example, there are baryons inside the dark matter halos in the form of gas and stars 
and these can redistribute the matter inside halos 
beyond what the N-body simulations can predict. Baryon gas has pressure and this already 
changes the total matter profiles significantly, and even more dramatic effects arise from some feedback models
where gas is pushed out of the halo center, possibly even dragging dark matter along \cite{2011MNRAS.415.3649V}. 
These processes will change parameters associated with the halo profile, such as $R_{nh}$ and $R_n$, 
(models of \cite{2011MNRAS.415.3649V} 
suggest these coefficients change at a level of 5-10\% \cite{2014MNRAS.445.3382M}), 
but because of mass conservation $A_0$ changes a lot less \cite{2014MNRAS.445.3382M}. 
Moreover, most of the cosmological information content is already in Zeldovich term and $A_0$ \cite{2014MNRAS.445.3382M}. 
So while HZPT, and PT in general, may only be able to determine two numbers beyond Zeldovich approximation, 
this may also be all that can be reliably extracted from N-body simulations at a two point function level. 

\section{Acknowledgements}
U.S. is supported in part by the NASA ATP grant NNX12AG71G. 
Z.V. is supported in part by the U.S. Department of Energy contract to SLAC no. DE-AC02-76SF00515.
We acknowledge useful discussions with T. Baldauf, L. Senatore, M. White and M. Zaldarriaga and we thank T. Baldauf for simulations data. 

\bibliography{cosmo,cosmo_preprints}
\end{document}